\documentclass[10pt,letterpaper,twocolumn]{article} 
\usepackage{ol2}
\usepackage[draft]{hyperref}
\usepackage{amsmath,amssymb}
\usepackage{graphicx}

\begin{document}

\twocolumn[ 

\title{Multiscale analysis of sub-wavelength imaging with metal-dielectric multilayers}

\author{Rafa{\l} Koty{\'n}ski, Tomasz Stefaniuk}
\affiliation{Department of Physics, University of Warsaw\\
Pasteura 7, 02-093 Warsaw, Poland\\
rafal.kotynski@fuw.edu.pl, tstefaniuk@igf.fuw.edu.pl
}

\begin{abstract}
Imaging with a layered superlens is a spatial filtering operation characterized by the point spread function (PSF).   We show that in the same optical system the image of a narrow sub-wavelength Gaussian incident field may be surprisingly dissimilar to the PSF, and the width of PSF is not a straightforward measure of resolution. FWHM or std. dev. of PSF give ambiguous information about the actual resolution, and imaging of objects smaller than the FWHM of PSF is possible. A multiscale analysis of imaging gives good insight into the peculiar scale-dependent properties of sub-wavelength imaging.
\end{abstract}
] 

\begin{figure}
\centering
\includegraphics[width=3.5in]{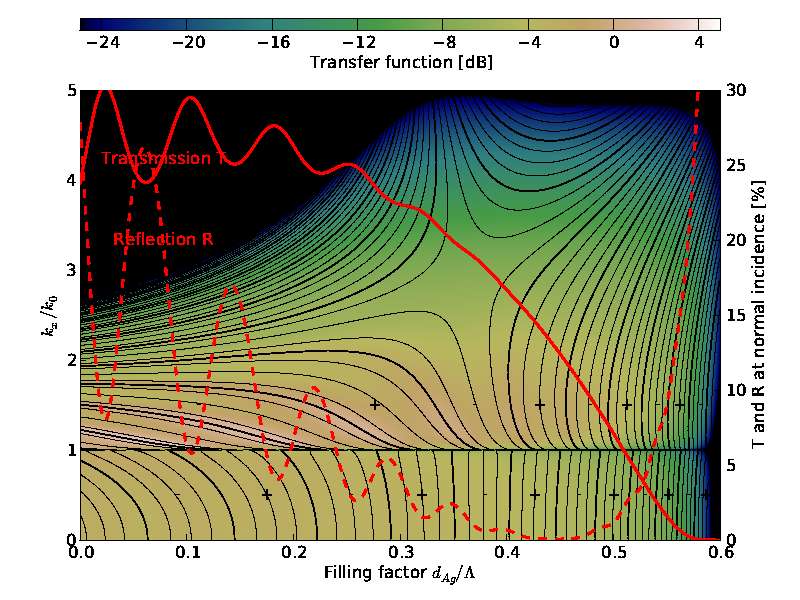}
\caption{Intensity transmission and reflection coefficients ($T,R$) of the multilayer as a function of the filling factor, plotted over the transfer function (in vertical cross-sections of the color map). Phase isolines are distanced by $\pi/4$.
}
\label{fig.tf}
\end{figure}

\begin{figure}
\centering
\includegraphics[width=3.5in]{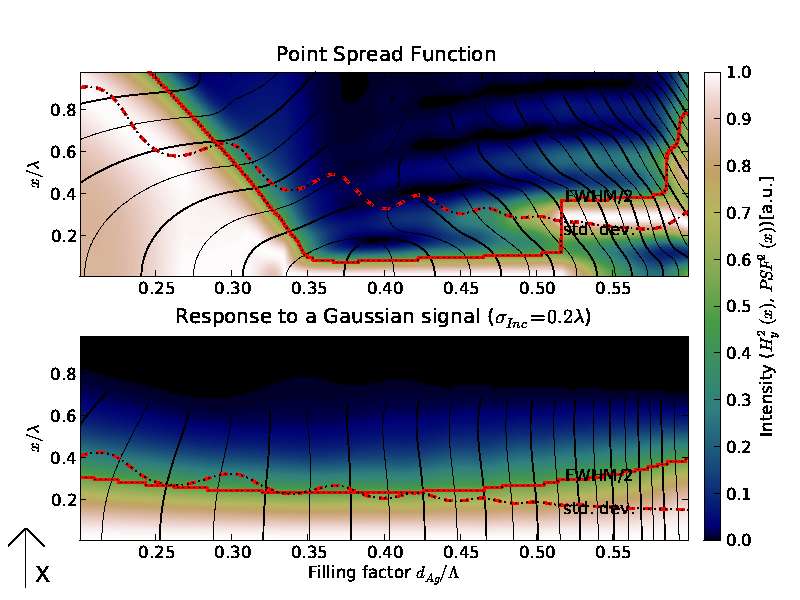}
\caption{Point spread function  (in vertical cross-sections) in the function of the filling factor (top) compared with the response of the multilayer to a narrow Gaussian signal with sub-wavelength width ($\sigma(H_y^2)=0.2\lambda$) (bottom). Phase isolines are distanced by $\pi/2$. }
\label{fig.psf}
\end{figure}

\noindent\label{sect:intro}
Since the seminal paper by Pendry\cite{pendry2000nrm} and the first experimental demonstration of sub-wavelength imaging through a $40nm$ silver slab\cite{fang2005sdl,melville2005sri}, sub-wavelength imaging at visible wavelengths has been investigated in much thicker low-loss layered silver-dielectric structures\cite{wood2006,Scalora2007opex,Scalora08pra,belov2005csi,belov2006prb,Li:prb07,Kotynski:jopa2009}. A variety of physical models may be applied to explain the mechanism of transmission: 1. the effective medium anisotropic approximation of the sub-wavelength multilayer\cite{wood2006} combined with the Fabry-Perot resonant condition tuned to be independent of the angle of incidence\cite{belov2006prb,Li:prb07}; 2. negative refraction at the boundary of silver slab\cite{pendry2000nrm}  multiple interfaces of the layered structure resulting in diffraction-free propagation\cite{Scalora08pra}; 3. resonant tunneling through the bandgap material formed by the periodic metal-dielectric multilayer\cite{Scalora2007opex}. Enhancement of evanescent fields needed for sub-wavelength imaging may be also explained in various ways: 1. as the result of collective coupling between plasmon modes at subsequent metallic layers\cite{Conforti:ol2008,Scalora:arxiv09} - if we look to the internal field distributions; 2. as self-collimation\cite{Kosaka:apl99} - if we look to the band diagrams of the multilayer; 3. as the result of large effective permittivity $\epsilon_{\bot}\approx \infty$ -  when the homogenized anisotropic model of the system is valid.

Here, we 
 calculate numerically the transfer function (TF) of the multilayer system. Let us  emphasize that the structures analyzed are  inaccurately characterized with the homogenization model which underestimates the value of transmission.

Imaging through a layered superlens consisting of uniform and isotropic materials is a linear and shift invariant (LSI) spatial filtering operation. Furthermore, planar imaging is a scalar LSI system. The theory of LSI systems, widely utilized in the diffraction theory and in Fourier Optics is therefore readily applicable. However, for systems with sub-wavelength resolution, point spread function (PSF) engineering leads to  surprising peculiarities. In this paper we show that the width of PSF is not a straightforward measure of resolution. 

\begin{figure}
\centering
\includegraphics[width=3.5in]{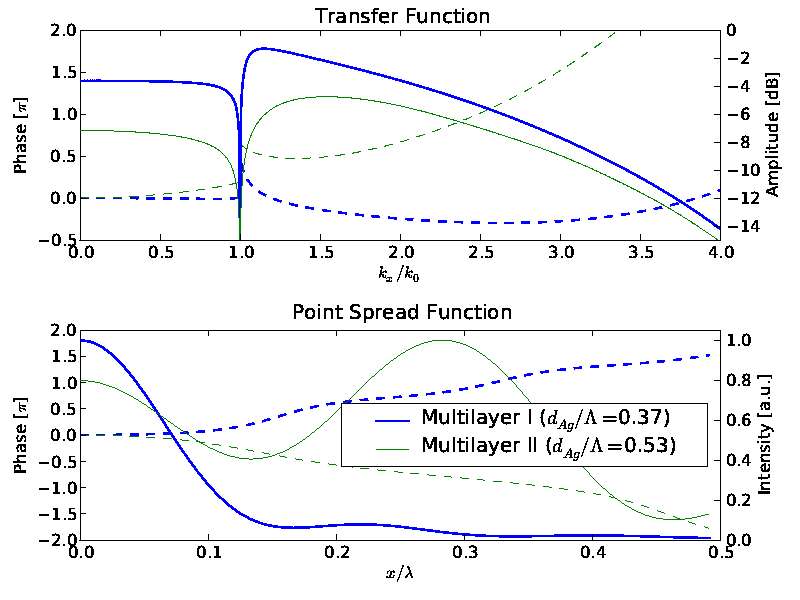}
\caption{TF (top) and PSF (bottom) of the two multilayers (solid line- amplitude/intensity; dashed line -phase).}
\label{fig.mtf1d}
\end{figure}

\begin{figure}
\centering
\includegraphics[width=3.5in]{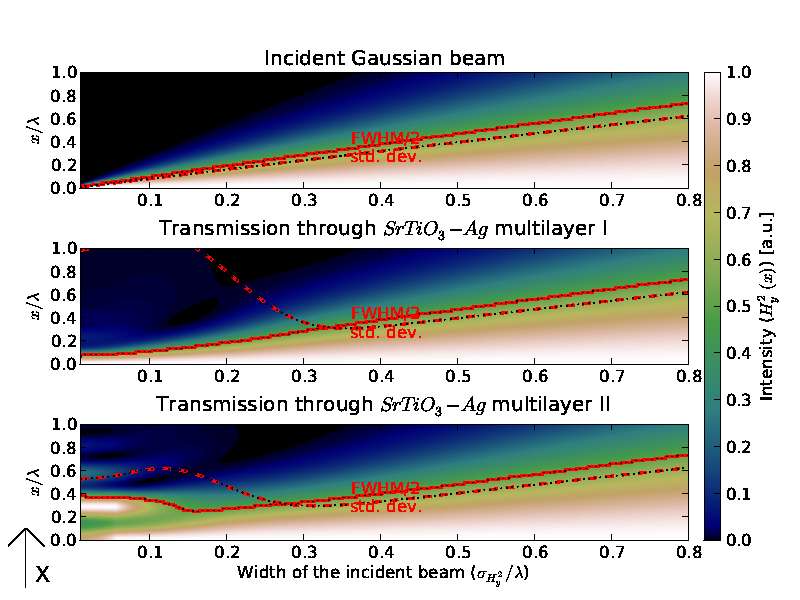}
\caption{Transmission of a sub-wavelength Gaussian incident field of various width through two multilayers (vertical cross-sections include the normalized field $H_y^2$).}
\label{fig.comp}
\end{figure}

For the TM polarization, the magnetic field $H_y(x,z)$ may be represented with its spatial spectrum $\hat H_y(k_x,z)$
\begin{equation}
 H_y(x,z)=\int_{-\infty}^{+\infty} \hat H_y(k_x,z) exp(\imath k_x x) dk_x,
\end{equation} 
where, at least for lossless materials, the spatial spectrum is clearly separated into the propagating part $k_x^2<k_0^2 \epsilon$ and evanescent part $k_x^2>k_0^2 \epsilon$. 

The TF is the ratio of the spatial spectra of the output and incident fields and corresponds to the amplitude transmission coefficient of the multilayer
\begin{equation}
\hat H_y(k_x,z=L)=TF(k_x) \cdot \hat H_y^{Inc}(k_x,z=0).
\label{eq.mtfdef}
\end{equation}
Due to reflections, the incident field $\hat H_y^{Inc}(k_x,z=0)$ differs from the total field $\hat H_y(k_x,z=0)$.

 PSF of an imaging system can be often straightforwardly interpreted, and provides clear information about the resolution, loss or enhancement of contrast, as well as the characteristics of image distortions. For instance, the resolution may be usually linked to the width of PSF. When the input signal and PSF are positive functions with a limited region of support (region with non-zero values), the regions of support of convolved functions simply add together, contributing to the broadening of the filtered signal. This can be expressed more conveniently using the following relation on the $L_0$ norms,
\begin{equation}
\|H_y(x)\ast PSF(x)\|_0=\|H_y(x)\|_0+\|PSF(x)\|_0,
\label{eq.norm0}
\end{equation}
where $\|f(x)\|_0= lim_{p\rightarrow 0^{+}} \int \lvert f(x)\rvert^p dx$.

On the other hand, for simple Gaussian PSF and input, the output has the width (variance) equal to the sum of variances of PSF and input,
\begin{equation}
exp(-x^2/\sigma_1^2) \ast  exp(-x^2/\sigma_2^2) \propto exp(-x^2/(\sigma_1^2+\sigma_2^2)).\label{eq.sgm}
\end{equation} 
Therefore once again the width of PSF has a clear link to the resolution of the imaging system. However, formulas~(\ref{eq.sgm}) and~(\ref{eq.norm0}) are often invalid for diffractive systems with complex PSF. 

From now on, we focus on a $SrTiO_3-Ag$ multilayer with $N=20$ periods, and the total thickness $L=1.15\mu m$. The elementary cell consists of an $Ag$ layer symmetrically coated with $SrTiO_3$. Strontium Titanate is an isotropic material with a high refractive index $n=2.674+0.027i$ at the wavelength $\lambda=430nm$\cite{Palik}. The refractive index of silver at the same wavelength is equal to $n_{Ag}=0.04+2.46i$\cite{JohnsonChristy}.
Fig.~\ref{fig.tf} shows how the TF of the multilayer depends on the silver-filling fraction, and the corresponding PSF is shown in the upper part of Fig~\ref{fig.psf}.  The evanescent part of the TF has a large magnitude, which is the necessary condition for sub-wavelength imaging.  The shape of TF is generally regular with the exception of the phase discontinuity in the vicinity of $k_x/k_0=1$, as well as  the strong phase modulation below $d_{Ag}/\Lambda\lesssim 0.35$ which suppresses the super-resolving properties of the PSF in that range. The phase step at $k_x/k_0=1$ in the TF influences the shape of the corresponding PSF which, with the increase of filling-factor, evolves from a narrow sub-wavelength maximum to a shape dominated by the side-lobe. The response to a narrow sub-wavelength Gaussian signal is entirely different from the PSF (Fig\ref{fig.psf}, bottom).
 PSF does not resemble a Gaussian function and its width measured with $FWHM$ is different from the doubled standard deviation. The off-axis background of PSF results in the high value of std. dev., and probably $FWHM$ is a more meaningful measure of resolution of the system. Moreover, the broadening of the optical signal can not be expressed with formulas~(\ref{eq.sgm}) or~(\ref{eq.norm0}). The exception is the range of filling fraction in between $0.35$ and $0.45$, where the PSF resembles a Gaussian function and the broadening follows a simple intuitive behavior.

More in general, the width of response may even show an anomalous (decreasing) dependence on the size of the sub-wavelength Gaussian incident signal. It is striking how dissimilar are the PSF and the response to a narrow Gaussian signal around $d_{Ag}/\Lambda\gtrsim 0.5$ (Fig~\ref{fig.psf}). The explanation is nevertheless not difficult, as the bandwidth of the TF in use depends (inversely) on the width of the incident Gaussian function. The opposite phase of TF for propagating and evanescent waves is the source of the side-lobe of PSF. Partial removal of the central maximum in PSF (equal to the mean value of TF) occurs only  when the contribution from evanescent and propagating harmonics to the mean value compensate each other. Broader Gaussian incident fields limit the bandwidth in use, and suppress this sensitive condition. 

 In Figs.~\ref{fig.mtf1d},~\ref{fig.comp} we show the TF, and PSF of two selected multilayers with the filling fractions equal  to $0.53$, and $0.37$, as well as their response to a sub-wavelength Gaussian field distribution with $FWHM<1.6\lambda$. These two multilayers represent the situation of a regular nearly Gaussian PSF and a side-lobe dominated PSF, respectively. Both multilayers allow for imaging of sub-wavelength details, however their responses scale differently with the size of sub-wavelength object. We attribute their different behavior to the value of the phase shift between the propagating and evanescent part of the TF~(fig.~\ref{fig.mtf1d}). We have recently analyzed an analogous situation\cite{Kotynski:jopa2009}, however resulting from the different modulation depth of TF.
\begin{figure}
\centering
\includegraphics[width=3.5in]{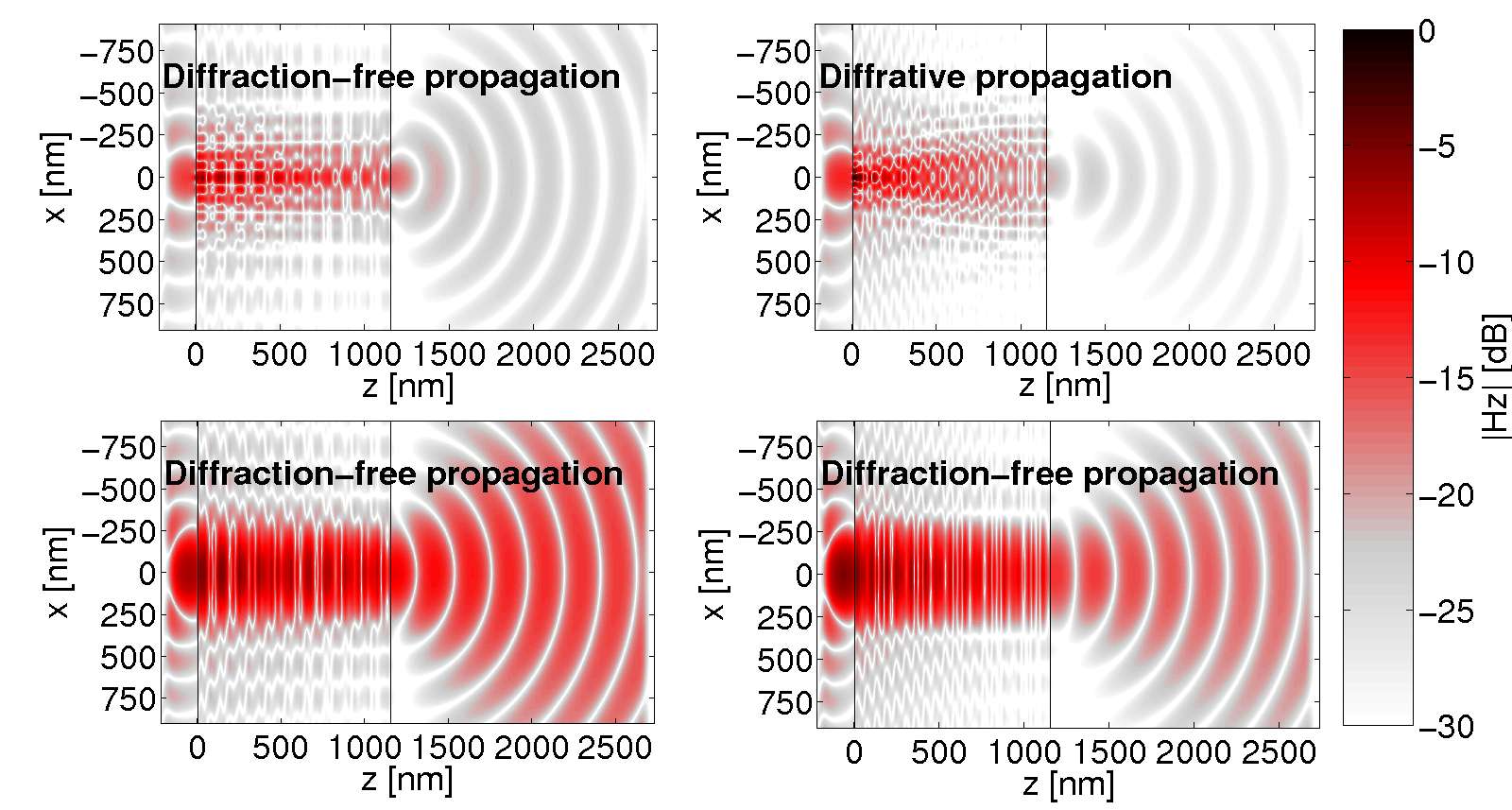}
\caption{Transmission of light through multilayer I (left) and multilayer II (right) (Instantaneous distribution of $\lvert H_y \rvert$ obtained with FDTD). The incident field is a point source (top) or a narrow Gaussian object with $\sigma(H_y^2)=0.2\lambda$ (bottom) }
\label{fig.fdtd}
\end{figure}

Finally, let us demonstrate the same example with FDTD simulation obtained with the open source Meep package~\cite{Farjadpour2006ol}. In Fig.~\ref{fig.fdtd} we show how a narrow Gaussian beam, as well as a beam originating from a point source propagate through the two discussed multilayers. The simulations only confirm the behavior described in Fig.~\ref{fig.comp} but are a good illustration to the peculiarities encountered in sub-wavelength imaging. While one of the multilayers allows for approximately diffraction-free propagation independently of the size of the source, the other behaves in the same way for broader sources only, and shows strong diffraction when the shape of the source approaches a $\delta$-function.

In conclusion, we have studied the transmission of sub-wavelength incident Gaussian field through a thick ($L\sim n \lambda$) silver-dielectric superlens. We have demonstrated that the response to narrow sub-wavelength Gaussian signal may be surprisingly different from the PSF of the system. Multiscale analysis provides the means to distinguish between diffraction-free propagation for various ranges of object sizes. 

We acknowledge support from the Polish projects \textit{N~N202~033237}, \textit{N~R15~0018~06}
 and the framework of COST actions \textit{MP0702}, \textit{MP0803}.


\end{document}